# Endeavors in the DualSPHysics code to evaluate heat transfer in nuclear systems

E. Mayoral-Villa[1], C. E. Alvarado-Rodríguez[2,3] *, F. Pahuamba-Valdez[4], J. Klapp[1], A. M. Gómez-Torres[1], E. Del Valle-Gallegos[4] and A. Gómez-Villanueva[1].

[1]*Instituto Nacional de Investigaciones Nucleares, Carretera México – Toluca s/n, La Marquesa, 52750 Ocoyoacac, Estado de México, México.*

[2]*Dirección de Cátedras CONACYT, Av. Insurgentes Sur 1582, Crédito Constructor, Benito Juárez, Ciudad de México 03940, México*

[3]*Departamento de Ingeniería Química, DCNyE, Universidad de Guanajuato, Noria Alta S/N, Guanajuato 36050, México*

[4]*Instituto Politécnico Nacional, Escuela Superior de Física y Matemáticas, Instituto Politécnico Nacional, Ciudad de México, México.*

*Corresponding author: ce.alvarado@ugto.mx

**Abstract**

This article condenses current endeavors and improvements in the expansion of applications of the DualSPHysics code to analyze heat transfer in a nuclear reactor core. This includes the essential conservation equations and certain physical considerations, particularly thermal conductivity variable model, considering changes in the reference density to maintain the accuracy in the solution. Conventionally, to study these sort of systems, Eulerian methods have been developed, nevertheless this kind of methods based on well-defined mesh show major restrictions. The state-of-the-art in nuclear safety is focused on Design Basis Accidents (DBA) and more recently also in Beyond Design Basis Accidents (BDBA). These analyses are associated with significantly more complex physical/chemical phenomena, involving a highly non-linear deformation. For these aims, an innovative approach is desirable for this kind of systems. The DualSPHysics code, based in Smoothed Particle Hydrodynamics (SPH) technique, has shown to be a real and robust alternative, since it involves a free mesh approach, and the numerical method is very well parallelized in both computational and graphical process units (CPU and GPU). The results for the improvements developed in the present work show an exceptionally good approximation with other simulation approaches and also with experimental observation in the three cases studied (1) heat transfer analysis in bidimensional system with thermal conductivity coefficient *k* variable, (2) natural convection heat transfer in a horizontal cylindrical ring similar to the space between the fuel rod and the cladding and (3) heat transfer in an experimental nuclear fuel rod square arrangement like in a Pressurized Water Reactor (PWR) nuclear core. Enhancements to this code (DualSPHysics) to use it in nuclear applications are fundamental in the exploitation of this technique in crucial areas of study.

# 1 INTRODUCTION

Study, research, and solution of problems associated with thermal hydraulics, neutronic and the dynamic of multiphase fluids involved in safety assessment, control and operation of a nuclear reactor using efficient and accurate numerical analysis is a mandatory necessity in the proposal of alternative environmental (clean air) energies such as nuclear energy.

The physical phenomena that take place during the operation of a nuclear reactor are extraordinarily complex. On one side, the neutronic properties of fuel strongly depends on the energy of the neutrons in charge of maintaining chain reaction, such energy, in turn, depends on the capacity of the coolant fluid to extract the heat and to moderate the energy of neutrons by means of scattering. When heat generated in fuel is transferred to the coolant fluid, changes in density and temperature are observed in the fluid and, in some cases, even phase changes. These changes in the coolant feedback in the neutronic properties because the original properties of fluid for cooling and moderating have changed due to the changes in density. In the case of Boiling Water Reactors (BWR), a transition from liquid to vapor is presented inside the core of the reactor. Thus, the properties of heat transfer are affected by the nucleation and the process of boiling alters the dynamics of the fluid and, in turn, the neutronics inside the fuel are also affected. Because of that, the transition of the liquid phase into gas plays a particularly important role in the design and control of a water reactor.

Based on the previously discussed, one of the most important thermal limits for design and operation of a nuclear reactor is the one related to the capability of the coolant to extract heat in an optimal way even in the extreme conditions that could be present in abnormal operation. A limit phenomenon in a BWR known as dry-out and in a PWR known as Departure from Nucleate Boiling (DNB), is related to the critical heat flux in the cladding of the fuel in which a film of vapor is formed in the cladding acting as a thermal isolator, damaging the heat transfer and thus, increasing dramatically the temperature in the fuel with the consequence of fuel overheats and melting.

The heat transfer in nuclear fuel takes part in several steps. The complexity of the heat transfer phenomena in the fuel and from fuel to coolant is increased by the nuclear reactions inside the fuel pellet during power generation. Nuclear reactions modify the original composition of the fuel pellet changing not only the chemical composition inside the solid fuel pellet but also delivering fission gases into the fuel gap, changing the composition of the gap and also the inner thermal and pressure loads.

On the other side of the cladding, the flow pattern of the coolant fluid is also an important issue in nuclear reactor analysis. The heat transfer to fluid strongly depends on the characteristics, behavior, and flow pattern of the coolant. For instance, the turbulent behavior of the fluid in a forced convection, favors the transfer of heat in the system.

Traditionally, to study these phenomena, Eulerian methods such as finite elements, finite volumes and finite differences have been developed. The codes based on Computational Fluid Dynamics (CFD) have demonstrated to be an important tool in the analysis of heat transfer phenomenon and, for some applications (heat transfer without boiling for instance) have been verified and validated. However, for the complex phenomenon of phase change, due to its chaotic behavior, Eulerian methods based on fixed mesh expose serious limitations.

To overcome these difficulties, the Smoothed Particle Hydrodynamics (SPH) method has been considered as a promissory option because it implies a free mesh methodology, and the numerical scheme is highly parallelized specially on the graphical process units (GPU).

The AZTLAN Platform project [1] is a Mexican national initiative which aims to have a platform for analysis and design of nuclear reactors. Among the codes belonging to the AZTLAN Platform (neutronics and thermalhydraulic codes), numerical simulations have been explored using the DualSPHysics [2] open-source code with own implementations to perform SPH parallel calculations for the heat transfer phenomena in nuclear reactors. As a first exercise, and with the objective to be able to analyze with detail the heat transfer in the fuel gap, the numerical method developed and implemented in DualSPHysics was verified and validated with experimental data reported in the literature showing that the implementation is suitable for the study of heat transfer in nuclear fuel reactors [3], further developments of own methodologies were published in [4]. In the worldwide, the use of SPH for nuclear thermohydraulic and safety applications has been recently discussed [5] and even demonstrated with the novel developed code SOPHIA [6]. The range of applications can be enormous, for example, some studies have been done for simulating breakups of jet pumps of nuclear reactors [7], propagation of waves for structural damage in case of tsunamis [8], and safety analysis related to simulation of molten core sloshing phenomena which can occur during a severe accident in nuclear reactors cooled by liquid metals [9].

In this paper, for illustration of its pertinence, three cases concerning with nuclear applications are analyzed: (1) heat transfer analysis in bidimensional system with thermal conductivity coefficient *k* variable, (2) natural convection heat transfer in a horizontal cylindrical ring similar to the space between the fuel rod and the cladding and (3) heat transfer in an experimental facility for nuclear fuel rod square arrangement like in a PWR core. Conclusions show that the enhancements to the DualSPHysics code to use it in nuclear applications are fundamental in the exploitation of this technique in crucial areas of study.

## 2 SPH METHODOLOGY

The Smooth Particle Hydrodynamics method (SPH) is a well-known Lagrangian technique, for this reason, a brief description about it concerning with this work is presented (for more details and applications the reader could consult references [10-14]). SPH method is founded by the interpolation integral theory, where the equations that govern the dynamics of the continuous media are transformed into integral equations using an interpolation function. The medium, that can be fluids and deformable solids, is represented by a finite set of observation points called "particles", by means of a smoothing procedure where the estimated value of a function *f(x)* at a point *x* is given by:

$$f(x) = \int_\Omega f(x')W(x - x', h)dx', \tag{1}$$

*x* is the position and *h* the smoothing length, which defines the domain of influence $\Omega$. The SPH method calculates discrete particle properties applying a smoothing kernel distribution function to consider the influence of surrounding particles. Even the characteristic features of the particle of interest are affected by all other particles in the global domain, usually only the effects of adjacent particles inside a smoothing radius designated as *2h* are included, and *h* describes a specified area above which the kernel will be nonzero. $W(x - x', h)$ is the smoothing function (or kernel) and is a function of the position *r* and the smoothing length *h*. Traditionally, the smoothing length *h* is kept constant, but in many cases, it is necessary to consider variable *h* value for each particle. In this work the use of *h* variable was implemented [15, 16].

Applying the interpolation integral (eq. 1), for the density of a fluid *ρ(x)*:

$$\int_\Omega \left[\frac{f(x')}{\rho(x')}\right] W(x - x', h)\rho(x')dx'. \tag{2}$$

The domain $\Omega$ is subdivided into *N* elements of volume or "particles" each one having mass $m_b$ and density $\rho_b$. The sum of the masses of all the particles gives the total mass of the fluid and the mass of each particle ($m_b$) is $m_b = \rho(x')dx'$, where *ρ(x')* is the density and *dx'* is the volume differential.

In the present work the conservation equations of momentum, mass, and energy in Lagrangian formalism are considered:

$$\frac{d\rho}{dt} = -\rho \nabla \cdot \boldsymbol{v}, \tag{3}$$

$$\frac{d\boldsymbol{v}}{dt} = \frac{-1}{\rho}\nabla P + \frac{\mu}{\rho}\nabla^2 \boldsymbol{v} + \boldsymbol{F}^B + \boldsymbol{g}, \tag{4}$$

$$\frac{dT}{dt} = \frac{1}{\rho C_p} \nabla \cdot (k\nabla T). \tag{5}$$

Where $\rho$ is the density, *t* is the time, *v* is the velocity vector, *P* is the pressure, $\mu$ is the viscosity, *T* is the temperature and, *g* is the gravitational acceleration. $\boldsymbol{F}^B$ is the buoyant force and is the responsible of the motion of the fluid due to the change in temperature, the Boussinesq approximation is used in this work expressed as:

$$\boldsymbol{F}^B = -\boldsymbol{g}\beta(T - T_r), \tag{6}$$

where, $\beta$ is the thermal coefficient of volumetric expansion, $T$ is the temperature of the fluid and $T_r$ is the reference temperature of the fluid.

In equation 5, $C_p$ is the heat capacity and, $k$ is the thermal conductivity coefficient. The value of the thermal conductivity coefficient per particle is calculated by the following expression:

$$k_a = \alpha \rho_f Cp. \tag{7}$$

where $\alpha$ is the thermal diffusivity coefficient.

With these considerations, the momentum, continuity, and energy equations can be discretized using the SPH formalism giving:

$$\frac{d\rho}{dt} = -\rho_a \sum_{b=1}^{N} m_b \frac{v_b}{\rho_b} \cdot \nabla_a W_{ab}, \tag{8}$$

$$\frac{dv_a}{dt} = -\sum_{b=1}^{N} m_b \left(\frac{P_a + P_b}{\rho_a \rho_b} + \Gamma\right) \nabla_a W_{ab} - \boldsymbol{g}\beta(T - T_r) + \boldsymbol{g}, \tag{9}$$

$$\frac{dT_a}{dt} = \frac{1}{Cp} \sum_{b=1}^{N} \frac{m_b(k_a + k_b)(r_a - r_b) \cdot \nabla_a W_{ab}}{\rho_a \rho_b (r_{ab}^2 + \eta)} (T_a - T_b). \tag{10}$$

Equations (8) - (10) are coupled by the Tait state equation:

$$P = B\left[\left(\frac{\rho}{\rho_r}\right)^\gamma - 1\right], \tag{11}$$

where P is the pressure, $\rho$ is the density of the fluid, $\rho_r$ is the reference density, $B = c_0^2 \rho_r/\gamma$, $\gamma = 7$ for liquids and $\gamma = 1.4$ for gases.

To consider the change in the reference density $\rho_r$ in equations 7 and 11 due to the temperature change, the following model is used by the coefficient of volumetric expansion:

$$V_f = V_0[1 + \beta(T_f - T_0)], \tag{12}$$

$V_f$ and $V_0$ are the final and initial volumes respectively and $T_f$ and $T_0$ are the final and initial temperature, respectively. Regarding density, mass, and volume ($\rho = m/V$) it is obtained:

$$\rho_f = \rho\left(\frac{1}{1 + \beta(T_f - T_0)}\right). \tag{13}$$

Following this expression, the reference density $\rho_r = \rho_f$ is evolved at each time step having a new density when the temperature changes. This calculation is performed per particle at each time step.

The time integration is performed using the Verlet algorithm provided by DualSPHysics, where the density, velocity, position, and temperature of particle *a* are increase from time $t^n$ to $t^{n-1} + \Delta t = t^n$ using the algorithm reported in [17]. No-slip boundary conditions are implemented at the walls using the method of dynamic boundary particles developed by Crespo et al. [18].

With the previous model the change of temperature and density of the fluid affects the coefficient of thermal conductivity considering a more robust model in comparison with the models that consider constant $k$ as it will be shown in the result's section.

# 3 MODELS

## 3.1. Model Case 1.

Case 1 is a two-dimensional system which consists of modeling the temperature profile $T(x, y, t)$ of a plate at certain moments of time $t$. The board has the following conditions at the border.

$$T(x, 0, t) = T(x, H, t) = T(0, y, t) = T(L, y, t) = T_1 \qquad (14)$$

The initial condition is $T(x, y, 0) = T_0$ where $T_0$ is an initial temperature uniform through the domain of Figure 3.1.1.

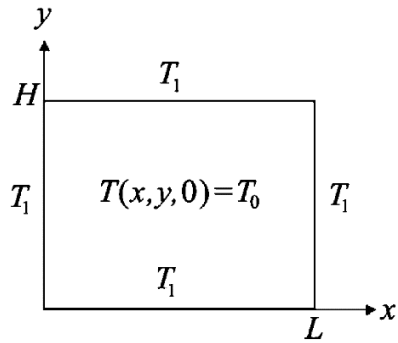

**Figure 3.1.1** 2D spatial domain with boundary conditions of constant temperature and initial conditions.

The governing equation for the development of the temperature field $T(x,y,t)$ for two-dimensional thermal conduction in Cartesian coordinates, is:

$$\frac{\partial T}{\partial t} = \alpha \cdot \left(\frac{\partial^2 T}{\partial x^2} + \frac{\partial^2 T}{\partial y^2}\right) \qquad (15)$$

where $\alpha = \frac{k}{\rho c_p}$ is the thermal diffusivity, $c_p$ is the specific heat, $k$ is the thermal conductivity and $\rho$ is the material density. Notice the dependence of the thermal conductivity with the density $\rho$.

The analytical solution of the temperature distribution field $T(x,y,t)$ can be determined in terms of an infinite double series [19].

$$T(x,y,t) = \frac{16T_0}{\pi^2} \sum_{k=1,3,...}^{\infty} \sum_{l=1,3,...}^{\infty} \frac{\exp\left[-\alpha_D \pi^2 \left(\frac{k^2}{L^2}+\frac{l^2}{H^2}\right)t\right]}{kl} \text{sen}\left(\frac{k\pi x}{L}\right) \text{sen}\left(\frac{l\pi y}{H}\right). \quad (16)$$

Figure 3.1.1 shows the problem to be solved, with the spatial domain established, the boundary conditions and the initial conditions. In this case, a square plate with dimensions $L = H = 10$ cm and boundary conditions of constant temperature, $T_1 = 0\,°$C is considered. The initial fluid temperature is $T_0 = 100\,°$C [19], and Figure 3.1.2 presents the spatial discretization of the 2D domain using N = 1,600 SPH fluid particles.

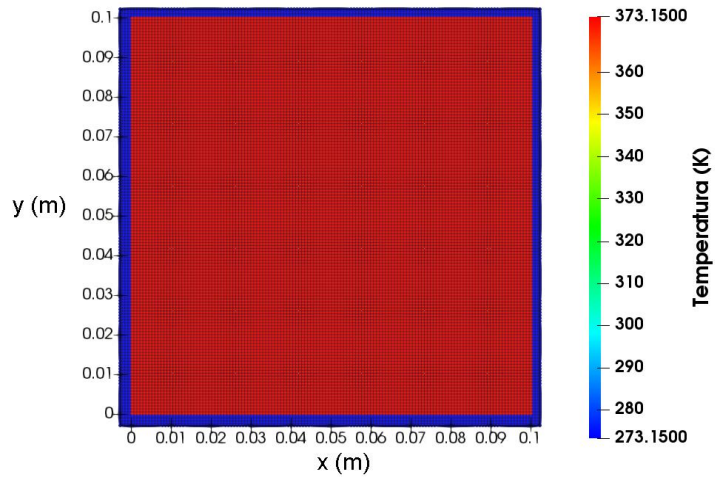

**Figure 3.1.2.** Spatial discretization of the 2D spatial domain using N = 1 600 SPH fluid particles.

Table 3.1.1 shows the parameters used in the simulation.

**Table 3.1.1** Simulation parameters used in Case 1.

| Parameter | Value |
| --- | --- |
| Initial distance between particles | 0.25 cm |
| Viscosity (Laminar Viscosity Treatment + SPS) | 1 x $10^{-6}$ m$^2$/s |
| Initial temperature of the fluid | 373.15 K |
| Step Algorithm | Verlet |
| Kernel | Wendland |
| Simulation time | 4 seconds |
| Temperature of the boundary | 273.15 K |
| Specific heat capacity at constant pressure | 4.1813 kJ/kg K |
| Thermal diffusivity coefficient | 1.0 x $10^{-4}$ m$^2$/s |
| Boundary particles | 81 |
| Fluid particles | 1600 |

## 3.2. Model Case 2

As it has been mentioned in the introduction, the heat transfer in nuclear fuel is a fundamental process involved in a nuclear reactor. The heat transfer in nuclear fuel takes part in several steps. First, the heat is generated in the fuel pellet by means of fission reaction and, it is transferred through it by heat conduction. Between fuel pellet and cladding, there is a space filled with an inert gas (usually helium) known as gap (see Figure 3.2.1). In the gap, heat transfers by convection from fuel wall to the inner cladding wall and further transfers through the cladding by conduction, until it reaches the coolant fluid on the outside wall of the cladding.

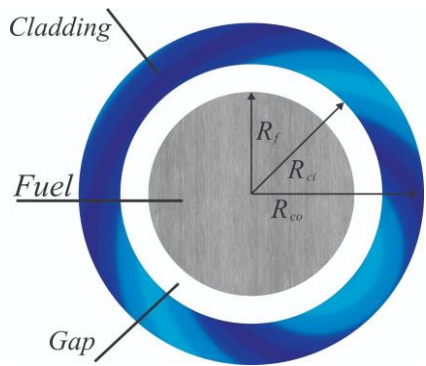

**Figure 3.2.1.** Structure of nuclear fuel.

This problem is like the natural convection heat transfer in a horizontal cylindrical ring, that has been a typical issue of study. The system can be described by two dimensionless parameters which are the Rayleigh number ($Ra$) and the Prandtl number ($Pr$). The Rayleigh number is defined as the ratio of the gravitational potential energy to the energy expected to viscous dissipation and thermal diffusion, and the Prandtl number is the ratio of viscous diffusivity to thermal diffusivity [20].

$$Ra = \frac{g \beta L^3 \Delta T}{\nu \alpha} \quad \text{and} \quad Pr = \frac{\nu}{\alpha}. \tag{17}$$

Where $g$ is the gravitational acceleration, $\beta$ is the thermal coefficient of volumetric expansion, $T$ is the temperature of the fluid, L the size, $\nu$ the viscous diffusivity and $\alpha$ the thermal diffusivity.

In general, the impact of $Pr$ number was normally not considered, nevertheless the numerical results indicate that the transition Rayleigh number from stable to unstable differs with Prandtl number [21]. Usually, the flow is stable at low Rayleigh number but unstable at high Rayleigh number and it has been found from numerical simulations [21] four different convection states explicitly: stable state with one plume, unstable state with one plume, stable state with multiple plumes, and unstable state with multiple plumes. It is thought that high thermal diffusivity (low $Pr$) enhances heat transfer, modifying the flow and inducing instability, however the flow states depend not only on the Rayleigh and Prandtl numbers but also on the vortex interactions [21]. In this sense, the SPH method can correctly replicate pure advection and, since in the SPH equations there are no explicit convection terms, additional computational resources for the convection terms are not necessary. These improvements are the motivations for the use of SPH method to explore natural convection in this work. In the SPH model used here, the body force is considered by the Boussinesq approximation (see equation 6) and it is originated by the adjustment of density in a temperature field.

The experimental system used for validating study Case 2 [22] is shown in the Figure 3.2.2, which consists of a setup of concentric cylinders supported horizontally. Both cylinders are at constant temperature, each one with different magnitude, the inner cylinder, as well as the components used to maintain the experimental conditions, have the highest temperature. The external diameter of the inner cylinder is 3.56 cm, with a thickness of 0.51 cm and the inner diameter of the outer cylinder is 9.25 cm, with a thickness of 0.45 cm, maintaining a relation $L/D_i = 0.8$.

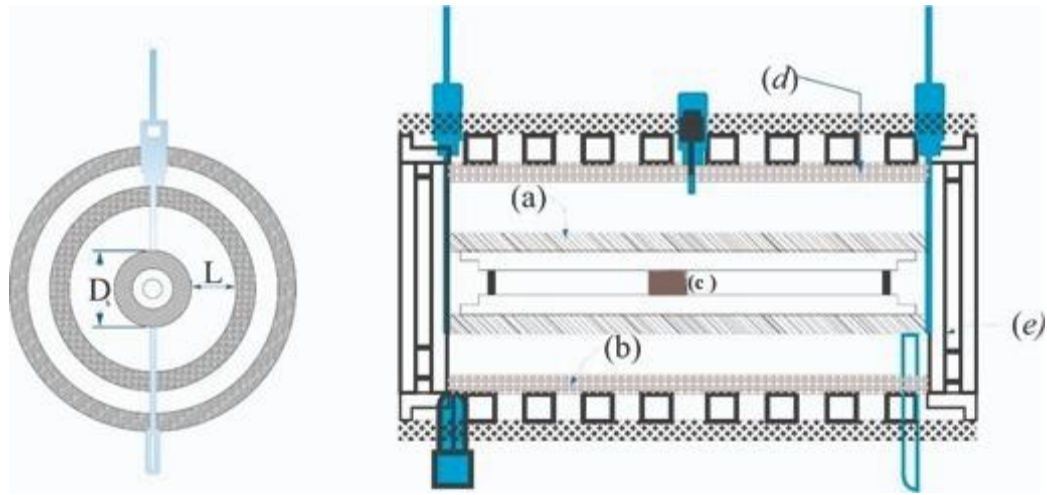

**Figure 3.2.2**. Illustration of the experimental setup used for validation [22]. (a) Internal cylinder. (b) External cylinder. (c) Heater. (d) Cooling water channels. (e) Window.

Following this array, for the mathematical modelling, two concentric tubes were simulated in which the same ratio of $L/D_i = 0.8$ was considered corresponding to Figure 3.2.3, where in (b) the initial conditions for the validation study cases are graphically represented. In all cases, the temperature of the borders is kept constant, and the only difference is the association that exists between the dimensionless numbers of Rayleigh ($Ra$) and Prandtl ($Pr$) giving by the value of the thermal diffusivity of the fluid ($\alpha$) as it is shown in equation 17.

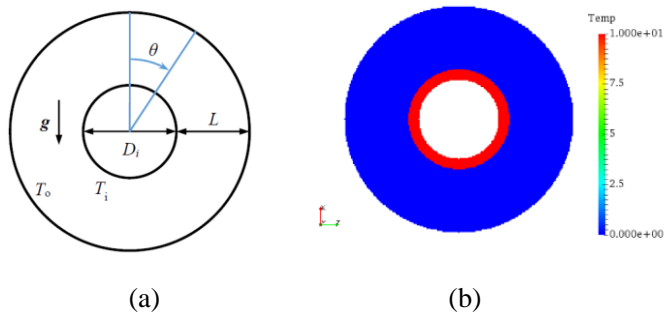

(a)                  (b)

**Figure 3.2.3** (a) Dimensions of the concentric tube system and (b) Initial conditions for SPH simulations

## 3.3 Model Case 3

For this case, a linear temperature distribution obtained from experimental data in the rods is studied using the setup described in Figure 3.3.1, which shows the experimental arrangement employed, corresponding with an experimental facility used for the measurement of voids of the central subchannel [23, 24], simulating one of the subchannels found in a PWR assembly type. Coolant flows axially into pressure vessel through the coolant inlet nozzle located just below the heated section. The effective heated length is 1555 mm, and the fraction measuring section of voids begins at 1400 mm from the bottom of the heated section [23]. In reference [24] different experimental measurements were done and reported, varying the input parameters, such as: flow of uniform heat over heated length, fluid inlet temperature, pressure, and mass flow rate. From this series of experimental measurements, the fraction of voids in the tests was calculated [24] and the results are indicated in the last column of Table 3.3.1.

**Table 3.3.1**: Initial conditions of the series of tests in steady state for measurement fraction of voids.

| Test number | Heat flux (W/cm$^2$) | Fluid inlet temperature $T_0$ (K) | P (MPa) | $\Delta T$ (K) | Mass flow rate (kg/s) | Voids fraction (%) |
|---|---|---|---|---|---|---|
| 1.2211 | 194.34 | 568.4 | 15.01 | 46.9 | 0.3248 | 3.8 |
| 1.2223 | 150.72 | 592.6 | 15.01 | 22.6 | 0.3248 | 31.1 |
| 1.2237 | 129.56 | 602.6 | 15.03 | 12.9 | 0.3248 | 44.0 |
| 1.4325 | 129.13 | 526.8 | 10.03 | 57.6 | 0.1487 | 33.5 |
| 1.4326 | 129.77 | 541.8 | 10.01 | 42.4 | 0.1487 | 53.1 |

**E. Mayoral et al.**

**Endeavors in the DualSPHysics code in nuclear systems**

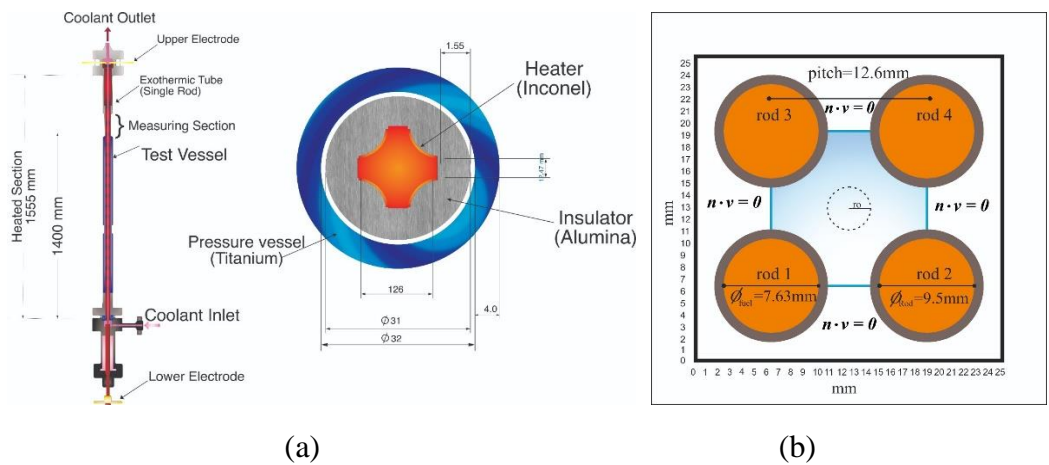

(a)                        (b)

**Figure 3.3.1**: (a) Test bench for the measurement of the central subchannel void distribution [23, 24]. (b) Cross section for void distribution measurement test of the central subchannel.

The representative subchannel for simulation is shown in Figure 3.3.1 (b), represented by the area with blue color and slanted lines (no slip boundary conditions), as well as the different dimensions: the diameter of the fuel (7.763 mm), the diameter of the fuel rod (9.5 mm), the distance between centers of the fuel rods known as pitch (12.6 mm) and the array size (25.2 mm × 25.2 mm).

From Table 3.3.1 it is observed that the test number 1.2211, presents the smaller voids fraction, for this reason and as a first step, these initial conditions were chosen to carry out the simulation in the current version of the DualSPHysics code. The other cases imply a significant generation of vapor (higher voids fractions) that, in turn, need a further implementation of change of phase capability in DualSPHysics code which is nowadays under development. Since experiment provides only heat flux from the heated perimeter, cladding temperature of the rods was obtained with the help of the sub-channel code SUBCHANFLOW [25] which receives as an input the heat flux in the fuel rod and delivers as an output the temperature distribution in the cladding outer wall. SUBCHANFLOW is a subchannel code with improved capabilities, where coolant properties and state functions are implemented for water using the IAPWS-97 formulation (The International Association for the Properties of Water and Steam), in addition, property functions for liquid metals (sodium and lead) are available, too. A detailed description of the code and its specific models can be found in [25-28].

The outer wall temperature of the cladding is used as a boundary condition in the DualSPHysics code. SUBCHANFLOW, being a mesh-based code, reports a temperature value in each simulated node, due to this, a fit was made to the values of the outlet temperatures to obtain a distribution function of the temperature in the cladding of a fuel rod. In this case, the 4 rods configuring the central subchannel have the same power distribution and therefore the same axial temperature profile shown in Figure 3.3.2.

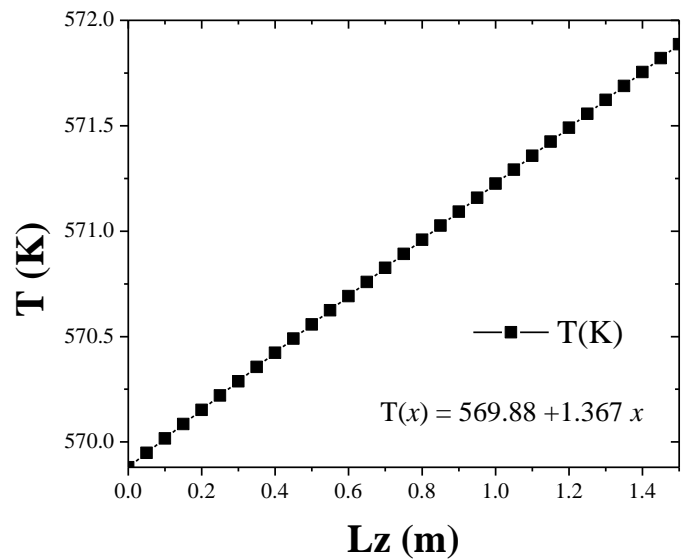

**Figure 3.3.2.** Lineal adjustment of the temperature values of the cladding.

Table 3.3.2 indicates the properties of water in the state of compressed liquid, required by the DualSPHysics code as input parameters, at a pressure of 15.01 MPa and at a temperature of 295.25 C.

**Table 3.3.2**: Properties of water at a Pressure of 15.01 MPa and a Temperature of 295.25 C.

| | |
|---|---|
| Density $\rho$ | 735.2184 kg/m$^3$ |
| Specific heat at constant pressure, $Cp$ | 5.3639 kJ/kg K |
| Sound velocity, $c$ | 992.8179 m/s |
| Kinematic viscosity. $\nu$ | 1.2274 X 10$^{-7}$ m$^2$/s |
| Thermic diffusivity, $\alpha$ | 1.4385 X 10$^{-7}$ m$^2$/s |
| Coefficient of volumetric expansion, $\beta$ | 2.6998 x 10$^{-3}$ K$^{-1}$ |

# 4 RESULTS

## 4.1. Case 1

Following the system described in section 3.1 results showed in Figure 4.1.2 were obtained. The isothermal plots calculated using the analytical solution of Eq. (16) and the SPH simulation results for the systems using conductivity model constant and conductivity model variable at t = 1 seg and 4 seg are shown in Figure 4.1.2. The SPH solution is symbolized by the points, and the analytical series solution by the solid lines.

The final test shows that results obtained with the SPH method presents good accuracy comparing with the analytical solution, especially in the case where the SPH variable thermal conductivity model is applied. Nevertheless, the stability of the implicit time integration schemes allowed greater time steps, while the solution accuracy was maintained [19], the solution was seen to break down as Δt was increased in the case of constant thermal conductivity. It is remarkable the improvement in the model using thermal conductivity variable considering changes in the reference density maintaining the accuracy in the solution even for large time steps comparing with the analytical result.

E. Mayoral et al.

Endeavors in the DualSPHysics code in nuclear systems

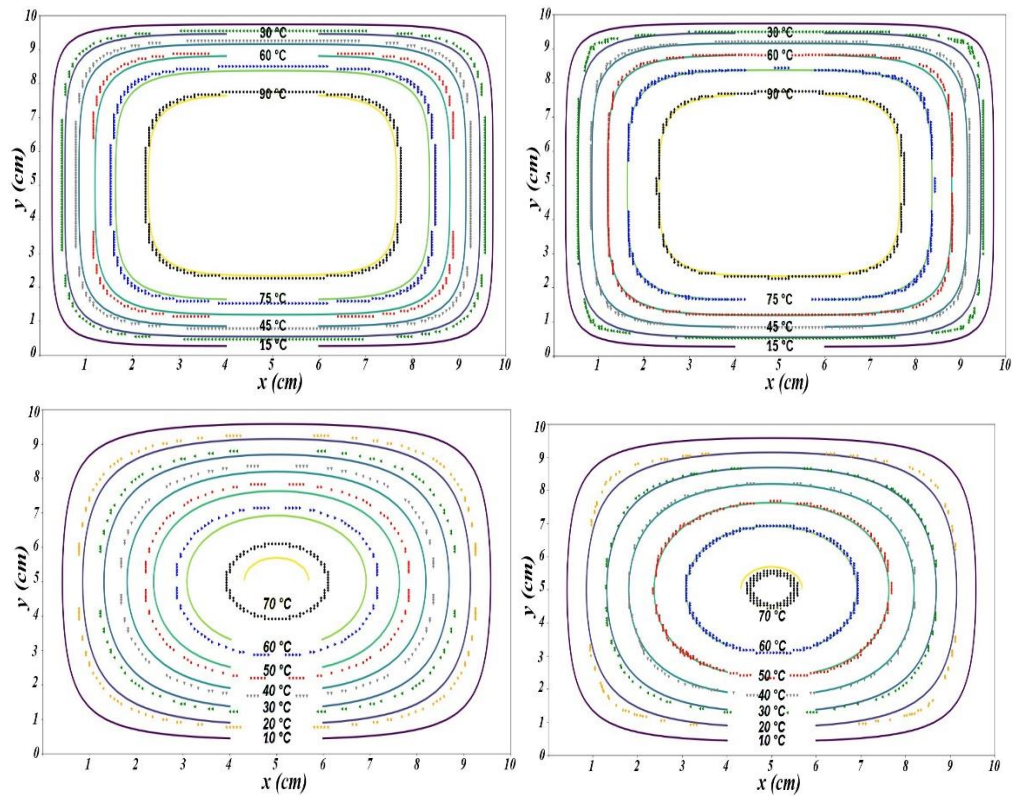

**Figure 4.1.1.** Temperature distribution (top) 1 seg (below) 4 seg. Conductivity model constant (left) and conductivity model variable (right)

### 4.2 Case 2.

Numerical simulations for different conditions for case 2, described in section 3.2, were performed using a fixed $Pr = 10$ and $Ra = 10^2$, $10^4$ and $10^6$. In all cases, the steady state was reached and then, the isothermal profiles between the concentric tubes were obtained. Figure 4.2.1 shows these results comparing them with the data reported in [21].

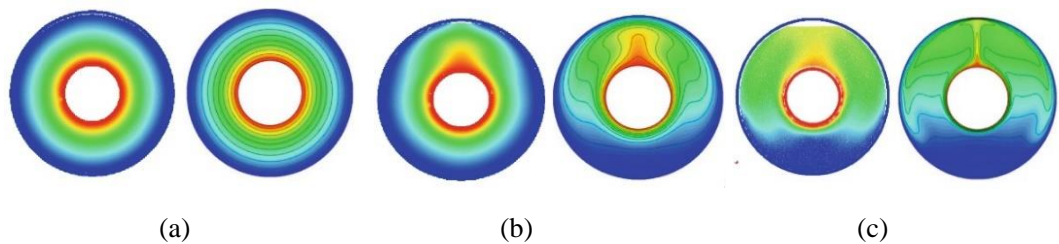

(a)          (b)          (c)

**Figure 4.2.1.** Comparison of the isothermal profile results with DualSPHysics (left) and those reported by Yang & Kong [21] (right), at $Pr =10$ and (a) $Ra = 10^2$, (b) $Ra = 10^4$ and (c) $Ra = 10^6$ respectively.

As expected, in the ring, the hot fluid will go up at the same time as the cold fluid goes down due to the buoyancy impact in a temperature field. As shown in Figure 4.2.1, the temperature of the fluid bordering on the hot interior limit is greater than that in the vicinity of the cold external border, producing the hot fluid to go up beside the central border and produce a plume at the center. The cold fluid goes down alongside the exterior frontier to the base of the ring. The natural convection of the fluid becomes greater, and the plume of the hot fluid becomes unstable when the Rayleigh number rises. In Figure 4.2.2 the (a) temperature distribution, (b) velocity profiles, (c) density variation and (d) change of thermal conductivity at different simulation times: 0 sec, 10 sec, 20 sec, 30 sec and 40 sec, for $Ra = 10^6$ and $Pr = 10$ are presented. Since Prandtl number is the ratio of viscous diffusivity to thermal diffusivity and due to the convection flow is driven by the buoyancy effect produced by thermal diffusivity, while the viscous diffusivity is trying to stop the convection, the states with more than one plume are not detected for high Prandtl numbers. This is observed in Figure 4.2.2. where at high $Pr = 10$, the viscosity is strong, and the convection is weak diminishing the possibility of creating vortices, for this reason only two convection states at high Prandtl numbers are observed.

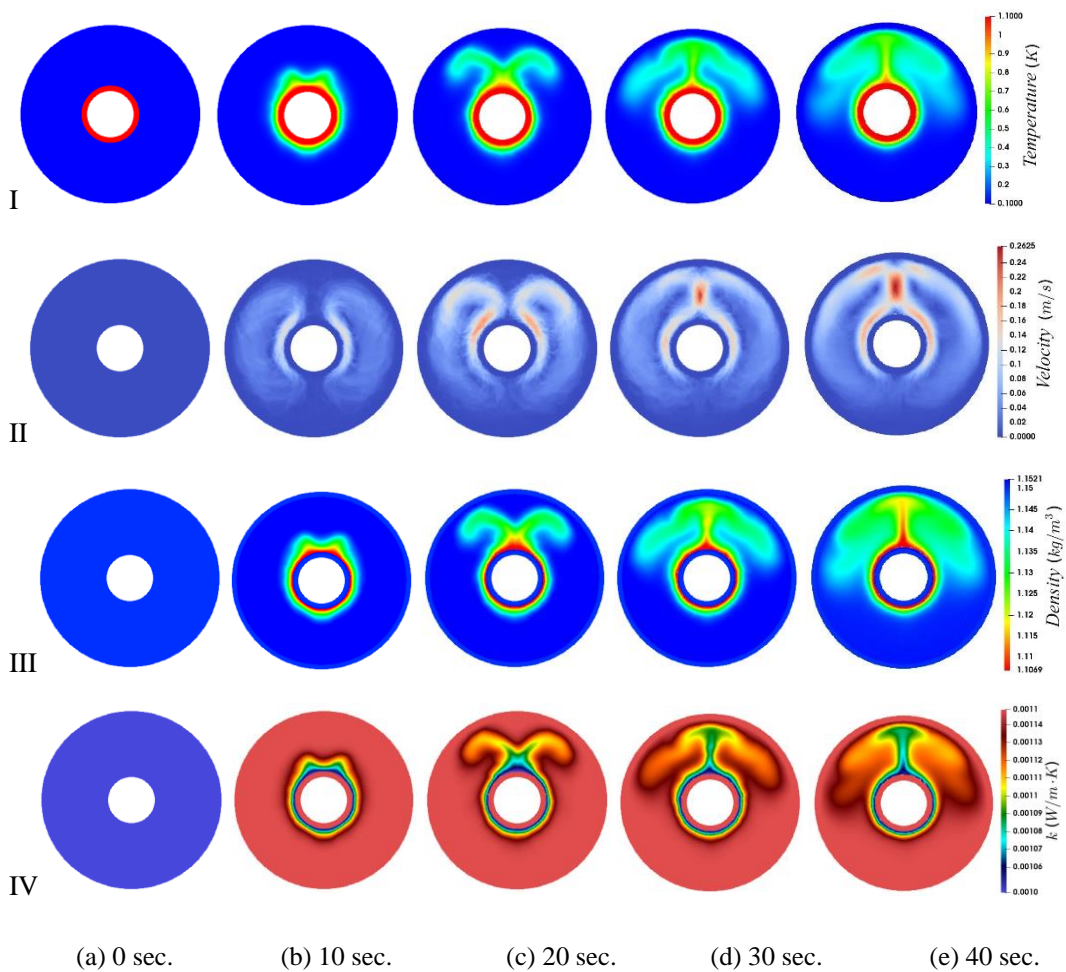

(a) 0 sec. (b) 10 sec. (c) 20 sec. (d) 30 sec. (e) 40 sec.

**Figure 4.2.2.** (I)Temperature (II) Velocity profiles (III) Density variation (IV) Change of thermal conductivity at different simulation times: (a) 0 sec. (b) 10 sec. (c) 20 sec. (d) 30 sec. (e) 40 sec.

A quantitative contrast between temperature profiles obtained with the DualSPHysics and experiments results reported in [22] is shown in Figure 4.2.3, where a graph of the cross section of a horizontal ring is displayed. R is the length from the center of the ring, and $\theta$ is the angle considered in clockwise direction from the vertical position (see Figure 3.2.3 (a)). $R_i$ and $R_o$ are the radii of the interior and exterior cylinders, correspondingly. $D_i = 2\,R_i$ is the diameter of the inner cylinder and $L = R_o - R_i$ is the space amongst the two concentric cylinders. $T_i$ and $T_o$ ( $T_i > T_o$ ) are the temperatures of the interior and exterior cylinders respectively. For comparison, the results making use of the non-dimensional temperature given by $(T - T_o)/(T_i - T_o)$ are presented.

As it can be seen in Figure 4.2.3., the performed calculations agree in a comprehensive manner with the experimental data.

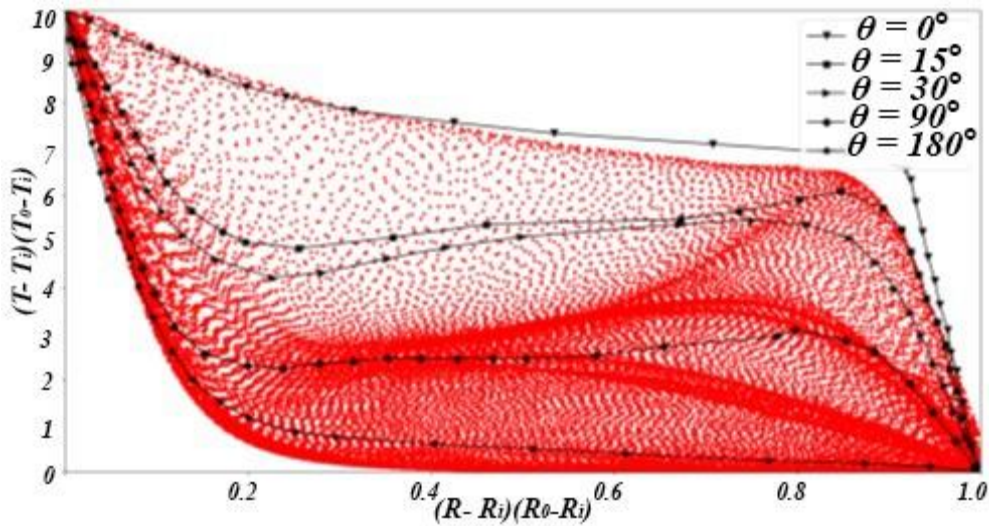

**Figure 4.2.3.** Dimensionless distribution of radial temperature in water for $Ra = 2.09 \times 10^5$, $Pr = 5.45$, $L/Di = 0.8$. The profiles for $\theta = 0°$, $15°$, $30°$, $90°$ and $180°$ were taken from [22] and compare with the SPH numerical results (red area).

### 4.3 Case 3

Using the setup described and presented in Figure 3.3.1 and the experimental linear temperature distribution showed in Figure 3.3.2, convergence analysis for the DualSHPysics system with $1 \times 10^6$, $2 \times 10^6$, $5 \times 10^6$ and $1 \times 10^7$ of fluid particles applying the conductivity model variable (VTCM) was performed. Results for this analysis are presented in Figure 4.3.1. showing that a good convergence is obtained using $1 \times 10^7$ particles of fluid. Simulation parameters used are presented in Table 4.3.1.

All convergence tests were performed in a computer with a GeForce RTX 2080 SUPER, compute capability of 7.5 and global memory of 8192 MB. The execution

times were $t_e$ = 819.99, 1977.21, 8734.68 and 17912.68 seconds for N = $1\times10^6$, $2\times10^6$, $5\times10^6$ and $1\times10^7$ particles, respectively.

**Table 4.3.1**: Simulation parameters used in Case 3 for convergence analysis using the DualSHPysics code with 1033280, 2052974, 5022720 and 10403328 of fluid particles applying the conductivity model variable (see text for details).

| Distance between particles in each case | 0.65, 0.525, 0.4, 0.31 mm |
|---|---|
| Artificial Viscosity | 0.1 |
| Initial fluid temperature | 294.85 K |
| Rod temperature distribution | $T(x) = 569.88 + 1.367x$ |
| Specific heat at constant pressure | 5.3639 kJ/kg K |
| Thermal diffusivity of fluid | $1.73 \times 10^{-7}$ m$^2$/s |
| Volumetric expansion coefficient | $2.6998 \times 10^{-5}$ °C$^{-1}$ |
| Speed of sound | 990 m/s |

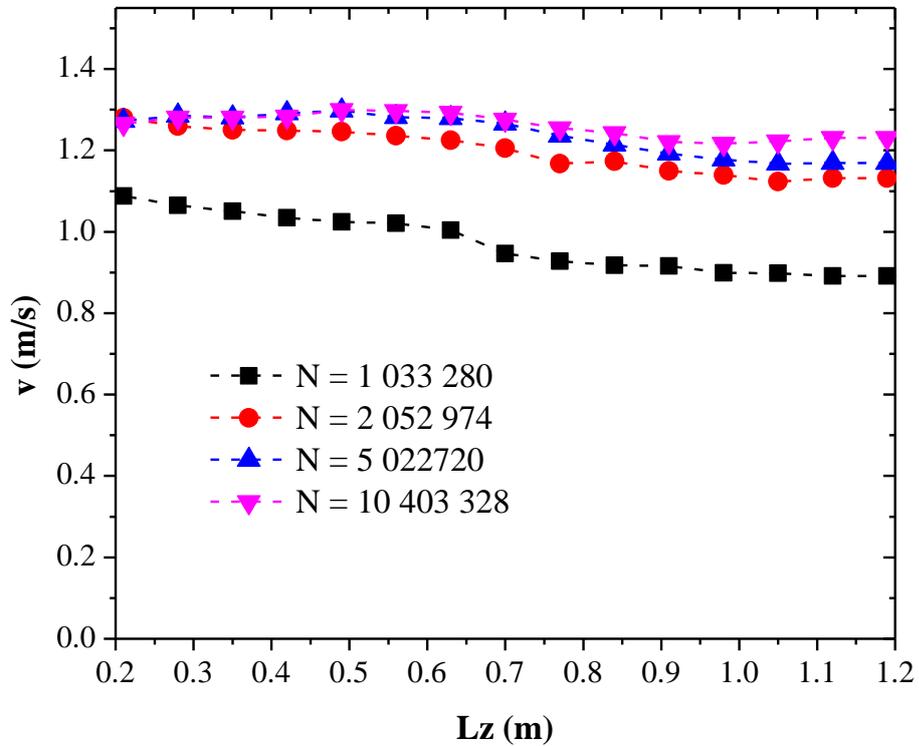

**Figure 4.3.1.** Convergence analysis for the system using 1033280, 2052974, 5022720 and 10403328 of fluid particles applying the conductivity model variable for Case 3 (see text for details).

Figure 4.3.2 (a) presents the velocity of the fluid for the case where the temperature distribution in the rods is linear, comparing the results obtained with our DualSPHysics code with 10 million particles of fluid and the SUBCHANFLOW code. The SPH results were achieved applying the variable thermal conductivity model (VTCM) and the constant thermal conductivity model (CTCM) using the simulation parameters presented in Table 4.3.1. In order to compare against SUBCHANFLOW, which produces punctual values representing the center of each axial node, the velocity shown in Figure 4.3.2 (a), is generated with DualSPHysics as a list of coordinates equivalent to a line passing through the center of the setup (Figure 3.3.1 (b)), then, the velocity is calculated at those coordinates. The velocity is interpolated from the particles closest to the coordinate set in the list.

As it is shown, the variable thermal conductivity model reproduces better the results obtained with the SUBCHANFLOW code for the velocity, meanwhile the constant conductivity model presents a lower value of the velocity. This result is expected because in the VTCM the buoyant force ($F^B$) is considered, and it is responsible of the motion of the fluid originated by change in temperature (see equation 6) for this reason a better approximation is obtained giving higher values than CTCM. In the case of SUBCHANFLOW, the heat transfer coefficient from cladding wall to fluid is determined by means of empirical correlations depending on the heat transfer mode. Some correlations have been verified and validated and are well documented in literature, however, some of them are restricted to operational limits and some other correlations are owner by fuel fabrication companies and inaccessible to researchers. Thus, the implemented VTCM in DualPhysics and verification and validation against SUBCHANFLOW (a code that is extensively used in nuclear field) implies a very good result that can overcome the need to have access to property correlations.

Additionally, fluctuation in the behavior of the velocity because of the temperature influence in the case of variable thermal conductivity model is also an indicative that more detail in the dynamics of the flow is captured with this model.

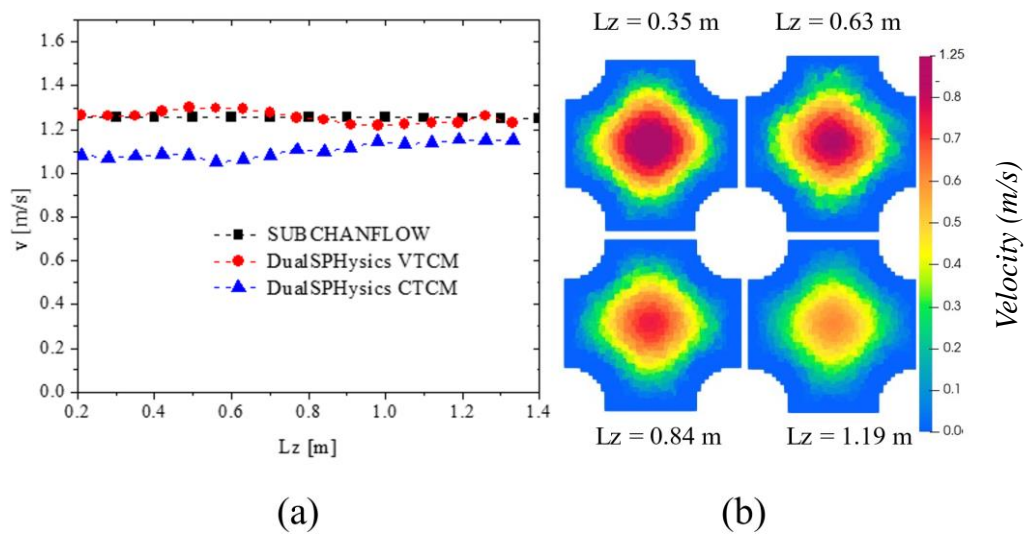

**Figure 4.3.2.** (a) Velocity of the fluid for the case where the temperature distribution in the rods is linear (b) velocity profiles in slices taken at different heights of the horizontal axes.

Figure 4.3.2 (b) presents the velocity fields in slices taken at different heights of the horizontal axes (0.35 m $\leq Lz \leq$ 1.19 m). Even though the velocity is almost constant, small fluctuations below average velocity can be observed when these profiles are analyzed, also it can be observed how less velocity is presented near to the borders; this is due to the no-slip boundary conditions that produces a diminish in the velocity in this area. Such kind of details cannot be captured by Eulerian codes like SUBCHANFLOW.

Figure 4.3.3. (a) presents the temperature profile for the simulations with DualSPHysics (VTCM and CTCM) comparing with the SUBCHANFLOW results. In the case of the results of the simulations with DualSPHysics for both VTCM and CTCM, the temperature value was obtained by means of the average of the temperature value of each of the particles around a certain radius $r_o$, whose origin is in the center of the system (see Figure 3.3.1 (b)). In Figure 4.3.3 (b), a detailed analysis of the temperature profiles in slices at different heights (0.8 m $\leq Lz \leq$ 1.06 m) of the rods is shown. As it can be seen, temperature distribution qualitatively corresponds well with the behavior expected, showing the radial increment in temperature due to the heat transferred from the rods to the fluid.

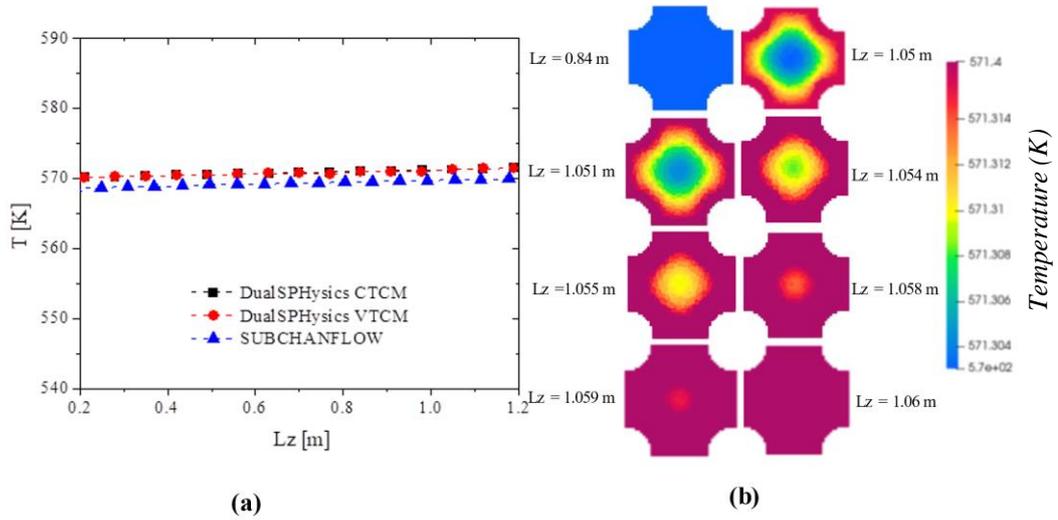

**Figure 4.3.3** (a) Temperature profile for the simulations with DualSPHysics (VTCM and CTCM) comparing with the SUBCHANFLOW results for Case 3 with linear temperature distribution in the rods. (b) Temperature profiles in slices at different heights (0.8 m $\leq Lz \leq$ 1.06 m) of the configuration.

Finally Figure 4.3.4 (a) presents the density ($\rho$) and (b) thermal conductivity constant ($k$) behavior of the fluid for case 3 using the DualSPHysics (VTCM and CTCM) comparing the results with the SUBCHANFLOW data. In this case, the density in DualSPHysics simulation is lower, and this corresponds to what it is expected since in the DualSPHysics simulation, the information of the change in density due to the variation of the volume given by the temperature is considered (see equation 12), therefore, a slight increase in the volume of fluid and therefore a decrease in its density is expected (see equation 13). The value of the thermal conductivity coefficient per particle also depends on changes in density (see equation 7), this is observed in the plot presented in Figure 4.3.4 (b), so more robust model in comparison with the traditional models that consider constant *k*, is obtained.

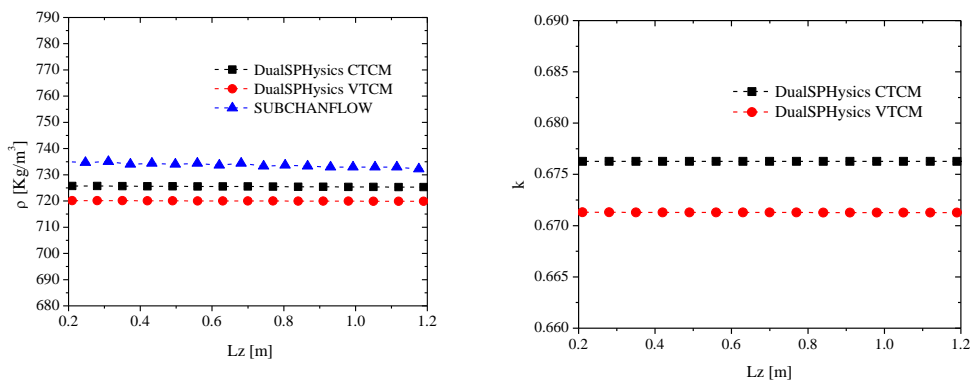

**Figure 4.3.4.** Density of the fluid and the *k* value for case 3 with linear distribution temperature using the DualSPHysics (VTCM and CTCM) comparing the results with the SUBCHANFLOW data.

# CONCLUSIONS

Improvements implemented in the DualSPHysics code for the study of heat transfer and its application in nuclear systems were demonstrated for three different applications. The developed model allows efficient and precise modeling of the studied phenomena, considering the variation in density and mass due to the increase in volume derived from the effect of temperature, allowing the distribution of speeds and temperatures of the system to be analyzed in greater detail. Results for the enhancements developed in the current work show a remarkably good approximation while comparing with other simulation approaches (SUBCHANFLOW) and also with experimental results in the three cases studied (1) heat transfer analysis in bidimensional system with thermal conductivity coefficient *k* variable, (2) natural convection heat transfer in a horizontal cylindrical ring similar to the space between the fuel rod and the cladding and (3) heat transfer in a nuclear fuel rod square arrangement like in a Pressurized Water Reactor (PWR). The most remarkable result is the implementation of the VTCM model which allows to have accurate results without depending on empirical correlations depending on the heat transfer mode. Some correlations are restricted to operational limits and some other correlations are owner by fuel fabrication companies and inaccessible to researchers. Enhancements to DualSPHysics presented in this work, for an extensive application in nuclear reactors heat transfer analysis, demonstrated the capabilities of the SPH approach to deal with typical phenomena in nuclear safety analysis. The exploitation of this technique will be crucial in this area of study and must be further developed, specially for leading with the phenomenon of change of phase in nuclear reactors which is one of the most important issues still being strongly investigated in the field.

# ACKNOWLEGMENTS

The authors acknowledge the financial support from the National Strategic Project No. 212602 (AZTLAN platform project) as part of the Sectorial Fund for Energy Sustainability CONACYT-SENER, and also the funding from the European Union's Horizon 2020 Program under the ENERXICO Project, grant agreement No. 828947 and under CONACYT-SENER-Hidrocarburos (Mexico), grant agreement No. B-S-69926. The calculations reported here were mostly performed using the supercomputing facilities of ABACUS Laboratorio de Matemática Aplicada y Cómputo de Alto Rendimiento of CINVESTAV-IPN.

**Conflict of interest**

**FIGURES**

**Figure 3.1.1.** 2D spatial domain with boundary conditions of constant temperature and initial conditions.

**Figure 3.1.2.** Spatial discretization of the 2D spatial domain using N = 1 600 SPH fluid particles.

**Figure 3.2.2.** (a) Dimensions of the concentric tube system and (b) Initial conditions for SPH simulations

**Figure 3.2.3**. Illustration of the experimental setup used for validation [22]. (a) Internal cylinder. (b) External cylinder. (c) Heater. (d) Cooling water channels. (e) Window.

**Figure 3.3.1**. (a) Test bench for the measurement of the subchannel void distribution central [23, 24]. (b) Cross section for void distribution measurement test of the central subchannel.

**Figure 3.3.2.**  Linear adjustment of the temperature values of the jacket.

**Figure 4.1.2.** Temperature distribution (top) 1 seg (down) 4 seg. Conductivity model constant (left) and conductivity model variable (right).

**Figure 4.2.1.** Comparison of the isothermal profile results with DualSPHysics (left) and those reported by Yang & Kong [21] (right), at Pr =10 and (a) Ra = $10^2$, (b) Ra = $10^4$ and (c) Ra = $10^6$ respectively.

**Figure 4.2.2.** (I)Temperature (II) Velocity profiles (III) Density variation (IV) Change of thermal conductivity at different simulation times: (a) 0 sec. (b) 10 sec. (c) 20 sec. (d) 30 sec. (e) 40 sec.

**Figure 4.2.3.** Dimensionless distribution of radial temperature in water for $Ra = 2.09 \times 10^5$, $Pr = 5.45$, $L/Di = 0.8$. The profiles for $\theta = 0°$, $15°$, $30°$, $90°$ and $180°$ were taken from [22] and compare with the SPH numerical results (red area).

**Figure 4.3.1.** Convergence analysis for the system using 1, 2, 5 and 10 million particles of fluid for Case 3 (see text for details)

**Figure 4.3.2** (a) Velocity profile for the simulations with DualSPHysics (VTCM and CTCM) comparing with the SUBCHANFLOW results for Case 3 with linear temperature distribution in the bars. (b) Velocity profiles in slices made at different heights (Lz) of the horizontal bars.

**Figure 4.3.3** (a) Temperature profile for the simulations with DualSPHysics (VTCM and CTCM) comparing with the SUBCHANFLOW results for Case 3 with linear temperature distribution in the bars. (b) Temperature profiles in slices made at different heights (Lz) of the horizontal bars.

**Figure 4.3.4.** Density of the fluid and the *k* for case 3 with linear distribution temperature using the DualSPHysics (VTCM and CTCM) comparing the results with the SUBCHANFLOW data.

TABLES

**Table 3.1.1** Simulation parameters used in Case 1.

**Table 3.3.1**: Initial conditions of the series of tests in steady state for measurement fraction of voids.

**Table 3.3.2**: Properties of water at a Pressure of 15.01 MPa and a Temperature of 295.25 C.

**Table 4.3.1**: Simulation parameters used in Case 3 for convergence analysis using the DualSHPysics code with 1033280, 2052974, 5022720 and 10403328 of fluid particles applying the conductivity model variable (see text for details).